\documentclass[10pt,aps,prb,twocolumn,showpacs,superscriptaddress]{revtex4-1}
\usepackage{hyperref}
\usepackage{graphicx}
\usepackage{epstopdf}

\usepackage[utf8]{inputenc}
\usepackage[english]{babel}

\usepackage[usenames]{color}
\usepackage{color}
\usepackage{colortbl}

\definecolor{linkcolor}{rgb}{0.9,0,0}
\definecolor{citecolor}{rgb}{0,0.6,0}
\definecolor{urlcolor}{rgb}{0,0,1}
\hypersetup{
    colorlinks, linkcolor={linkcolor},
    citecolor={citecolor}, urlcolor={urlcolor}
}

\begin{document}
\title{TUNABLE HYBRID TAMM-MICROCAVITY STATES}

\author{P. S. Pankin}\email{p.s.pankin@mail.ru}
\affiliation{Siberian Federal University, Institute of Engineering Physics and Radio Electronics, Krasnoyarsk, 660041, Russia}
\affiliation{Siberian Federal University, Institute of Nanotechnology, Spectroscopy and Quantum Chemistry, Krasnoyarsk, 660041, Russia}
\author{S. Ya. Vetrov}
\affiliation{Siberian Federal University, Institute of Engineering Physics and Radio Electronics, Krasnoyarsk, 660041, Russia}
\affiliation{Kirensky Institute of Physics, FRC KSC SB RAS, Krasnoyarsk, 660036, Russia}
\author{I. V. Timofeev}
\affiliation{Kirensky Institute of Physics, FRC KSC SB RAS, Krasnoyarsk, 660036, Russia}
\affiliation{Siberian Federal University, Institute of Nanotechnology, Spectroscopy and Quantum Chemistry, Krasnoyarsk, 660041, Russia}

\date{\today}

\begin{abstract} 
Spectral manifestations of hybrid Tamm-microcavity modes in a 1D photonic crystal bounded with a silver layer and containing a nematic liquid crystal layer working as a microcavity have been studied using numerical simulation.
It is demonstrated that the hybrid modes can be effectively tuned owing to the high sensitivity of the liquid crystal to the temperature and external electric field variations.
It is established that the effect of temperature on the transmission spectrum of the investigated structure is most pronounced at the point of the phase transition of the liquid crystal to the isotropic state, where the refractive index jump is observed.

\end{abstract}


\maketitle

\section{Introduction}

The optical Tamm state~\cite{Kavokin2005} is a surface state that is implemented when light is trapped between two mirrors.
This state can occur at the interface between a photonic-crystal (PC) Bragg mirror and the other PC~\cite{Kavokin2005a,Timofeev2016t}, left-handed medium~\cite{Namdar2006} or metallic mirror with negative permittivity $(\varepsilon<0)$~\cite{Kaliteevski2007,Sasin2008,Bikbaev2013mt}.
In the latter case, the light wave is merged with the surface plasmon, i.e., vibrations of free electrons at the surface of a metal and is called the Tamm plasmon polariton (TPP).
In contrast to the surface plasmon polariton, the TPP can be excited for both TM and TE light polarizations even under normal incidence of light onto the interface.

The TPPs have found application is lasers ~\cite{Symonds2012}, single-photon sources ~\cite{gazzano2012}, sensors and optical switches ~\cite{Afinogenov2016, Zhang2010_TammNL}, optical filters ~\cite{Pankin2016m_JOpt}, heat emitters ~\cite{ChenKuoping2016}, and nonlinear amplifiers~\cite{Vinogradov2006}.

The TPP mode can be hybridized with the modes of other types simultaneously excited in a system, e.g., with the exciton mode ~\cite{Kavokin2005, Symonds2009, Kaliteevski2011} or surface plasmon polariton ~\cite{Baryshev2012,Liu2012,Afinogenov2013}.
A new type of waveguide modes arises upon hybridization of two TPPs localized at the edges of a PC bounded with metallic or nanocomposite layers~\cite{Iorsh2012a,Bikbaev2013mt,Pankin2017}.

In recent years, close attention of researchers has been focused on the hybrid modes arising during simultaneous excitation of the TPP and microcavity (MC) mode in a PC system~\cite{Kaliteevskii2009}.
In studies~\cite{Bruckner2011, Bruckner2012}, the hybrid Tamm states were experimentally found.
The resonance wavelengths were tuned by changing the polarization of incident light and by designing a structure with the embedded metallic layer of variable thickness and its scanning by a small-aperture light beam.
The authors of~\cite{Zhang2013} proposed a new design of CuPc-PTCBI-based organic solar cells, where they implemented an idea of dual-states-induced broader-band absorption corresponding to the hybrid modes.
The idea of emitting light at two resonance wavelengths corresponding to the hybrid modes underlies the design of white top-emitting organic light-emitting devices (WOLEDs) based on the two-complementary-color strategy~\cite{Zhang2015}.
The experimental WOLEDs have the improved viewing characteristics and electroluminescence efficiency at the high quality of white color maintained.
The WOLEDs have a great potential of application in energy-efficient solid-state lighting sources and full-color flat-panel displays.
In addition, the formation of three hybrid modes upon simultaneous excitation of a pair of Tamm modes and one microcavity mode was investigated.
One of the hybrid modes ensures the extraordinary field amplification in a MC, which opens the way to intensification of the nonlinear optical effects~\cite{Fang2013}.
The hybridization makes it possible not only to enhance the field inside a MC, but also  to weaken absorption in the metallic layers embedded in the structure, which is important for designing vertical cavity lasers~\cite{Kaliteevski2015}.
In the above-cited works, the resonance wavelengths and light energy distribution over a structure could be tuned by selecting parameters of the structure during its fabrication only.

The authors of~\cite{Da2009, Luo2011} proposed to electrically control the Tamm state occurring at the interface between two PCs, one of which contains a nematic liquid crystal (nematic).
This can be realized owing to the high sensitivity of orientation in nematic layers to the applied voltage, which is reflected on the refractive indices of the layers and shifts the PC band gap and thereby the Tamm state wevelength.

The possibility of controlling MC modes in a 1D PC containing the nematic as a defect was demonstrated in~\cite{Arkhipkin2008,Zyryanov2010,Arkhipkin2011}
The MC modes were controlled using temperature, electric, and magnetic fields applied to the defect layer and by changing the angle of incidence and polarization of light.

In this work, we propose to tune the hybrid Tamm MC states via controlling the MC mode involved in the hybridization.
The MC mode is controlled via affecting the nematic defect in a 1D PC, which plays the role of a MC, by temperature or electric fields.
The PC is coated with a silver layer, which allows exciting the TPP at the PC/metal interface simultaneously with the MC mode on a defect.

\section{Model}
\begin{figure}[tbp]
\includegraphics[scale=1]{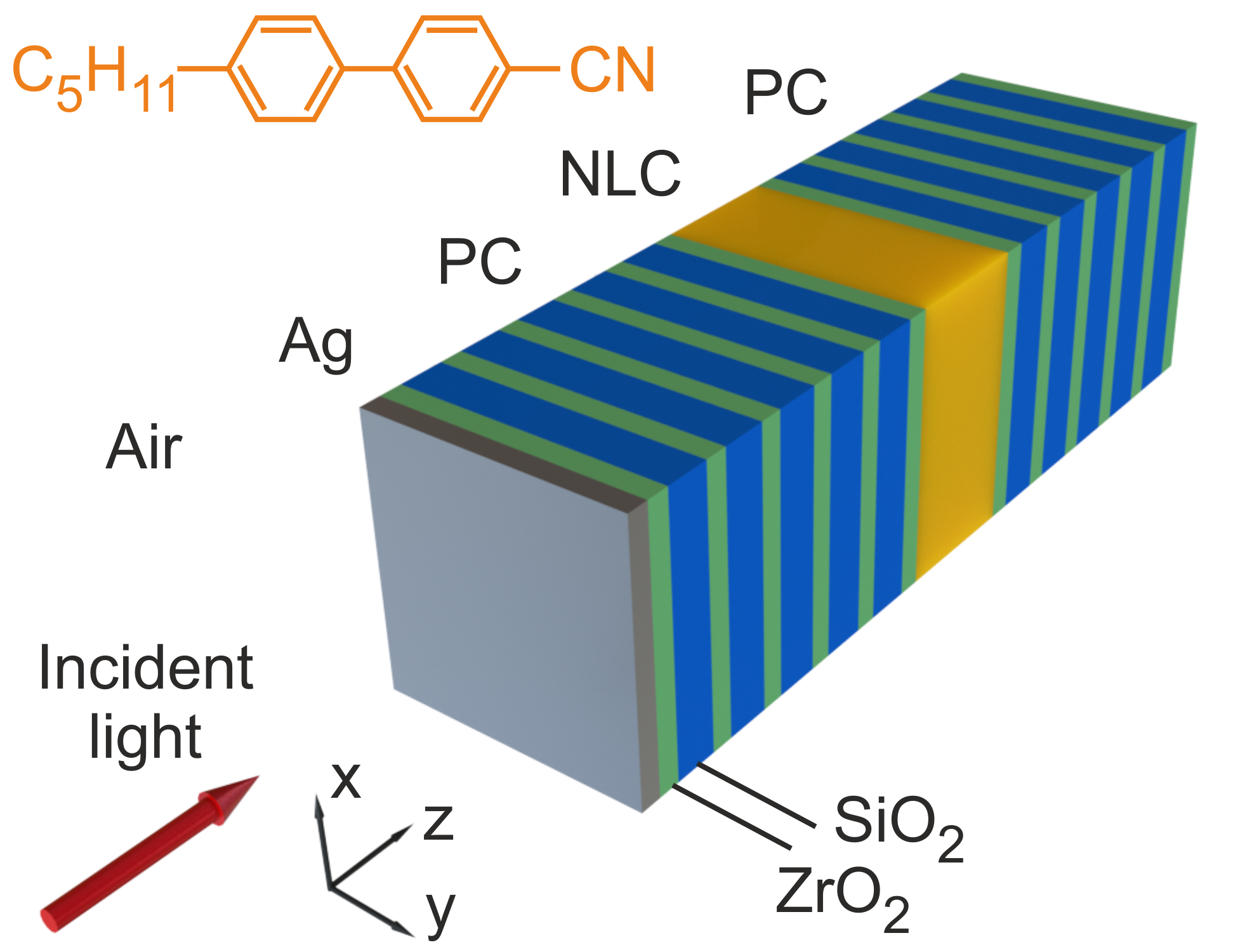}
\caption{Photonic crystal with a nematic defect bounded with a silver film.
Inset: structural chemical formula of the 5CB nematic.}
\label{fig1}
\end{figure}

The model under study is a 1D PC containing the nematic defect bounded with a silver film (Fig.~\ref{fig1}).
The PC unit cell consists of zirconium dioxide $ZrO_2$ and silicon dioxide $SiO_2$ with respective refractive indices and thicknesses of $n_a$ = 2.04, $W_a$ = 52~nm and $n_b$ = 1.45, $W_b = 102$~nm.
The defect layer with a thickness of $L = 2.13$ mkm is filled with a planar oriented 4-n-pentyl-4-cyanobiphenyl (5CB) nematic.
The nematic director, i.e., the unit vector of the preferred orientation of molecules, is aligned along the $x$ axis of the system.
The 5CB nematic undergoes a sequence of phase transitions: crystal -- $22.5^\circ$C -- nematic -- $35^\circ$C -- isotropic liquid.
The PC is coated with a silver film with a thickness of $W_{Ag} = 50$ nm on its one side; its permittivity is determined using the Drude--Sommerfeld approximation

\begin{equation}
\varepsilon _{Ag} \left( \omega \right)=\varepsilon _\infty -\frac{\omega _p^2 }{\omega \left( {\omega +i\gamma } \right)}.
\label{eq1}
\end{equation}
Here, $\varepsilon _\infty$ is the constant, which takes into account the contributions of interband transitions of bound electrons; $\omega _{p}$ is the plasma frequency; and $\gamma $ is the reciprocal electron relaxation time.
For silver, we have $\varepsilon _\infty$~=~5, $\omega _{p}$~=~9 eV, and $\gamma $~=~0,02~eV~\cite{Johnson_Christy1972}.
The total number of layers, including nematic and silver, is N = 24.

The transmission spectra of the structure and light field energy distribution in it were found numerically using the transfer matrix method ~\cite{Yeh1979} at the normal incidence of light.
We investigated the incidence of waves polarized along the $x$ and $y$ axes.

The LC structure in the applied voltage was calculated using the free energy variation technique \cite{Deuling1972}.
The elastic energy of the LC layer is

\begin{equation}
2F_k =(k_{11} \cos ^2\theta +k_{33} \sin ^2\theta) (d\theta /d z)^2.
\label{eq2}
\end{equation}
Here, $\theta$ is the tilt angle of the LC director relative to the $x$ axis and
coefficients of elasticity $k_{11} $ and $k_{33} $ correspond to the longitudinal bend and splay, 
respectively.
The electrostatic energy of the LC layer is expressed as

\begin{equation}
2F_e =-\vec {D}\vec {E}=-D_z^2 /\varepsilon _0\left( {\varepsilon _\bot \cos ^2\theta +\varepsilon _{\vert \vert } \sin ^2\theta } \right).
\label{eq3}
\end{equation}
Here, $\vec {E}$ is the vector of electric field applied to the LC layer, $\vec {D}$ is the vector of electric induction in the bulk of the LC, $\varepsilon _\bot$ and $\varepsilon_{\vert \vert } $ are the LC permittivities transverse and longitudinal relative to the director, and   
$\varepsilon _0$ is the permittivity of free space.

The free energy variation $F = F_k + F_e$ can be described as 
\begin{equation}
\delta F = \frac{\delta F_k}{\delta \theta} \;\delta \theta + \frac{\delta F_e}{\delta \theta} \;\delta \theta.
\label{eq4}
\end{equation}
Here, $\delta /\delta \theta$ is the operator of variational derivative with respect to orientation.
In the equilibrium configuration, free energy variation (\ref{eq4}) should be zero, regardless of the $\delta \theta$ value.
Taking into account formulas (\ref{eq2},\ref{eq3}), we arrive at the equation for the angle $\theta(z)$

\begin{eqnarray}
(k_{11} \cos ^2\theta +k_{33} \sin ^2\theta) \frac{d^2\theta} {dz^2}&+& \nonumber \\ 
\frac{(k_{33} - k_{11} )\sin (2\theta)}{2} \left(\frac{d\theta} {dz}\right)^2&+&
\nonumber \\
\frac{D_z^2 (\varepsilon _{\vert \vert } - \varepsilon _\bot) \sin (2\theta) }{2 \varepsilon _0 \left( {\varepsilon _\bot \cos ^2\theta +\varepsilon _{\vert \vert } \sin ^2\theta } \right)^2} &=& 0.
\label{eq5}
\end{eqnarray}

If the problem is one-dimensional and there is no divergence of induction ($\nabla \vec {D}=0$), we may conclude that this quantity is constant over the entire LC volume ($\left| {\vec {D}} \right|=D_z =const\left( z \right)$) and can be related to voltage $U$  applied to the defect layer as

\begin{equation}
U = \int_0^LE_zdz = \frac{D_z}{\varepsilon _0} \int_0^L \frac{dz}{\varepsilon _\bot \cos ^2\theta +\varepsilon _{\vert \vert } \sin ^2\theta}.
\label{eq6}
\end{equation}

The function of distribution of the tilt angle $\theta(z)$ is found from the joint solution of Eqs. (\ref{eq5},\ref{eq6}) with regard to the boundary conditions, which are determined by the rubbing direction on the PC mirrors.

\section{Results and discussion}

\subsection{Hybrid modes}

The formation of hybrid modes was demonstrated at the normal incidence of the $x$-polarized light wave, when the nematic director is parallel to the electric field.
The mode coupling phenomenon was investigated assuming the 5CB extraordinary and ordinary refractive indices $n_e = 1.71 + 0.00039i$ and $n_o = 1.54 + 0.00039i$ ~\cite{Li2005} with disregard of dispersion at a temperature of $25^\circ$C.

\begin{figure}[htbp]
\includegraphics[scale=1]{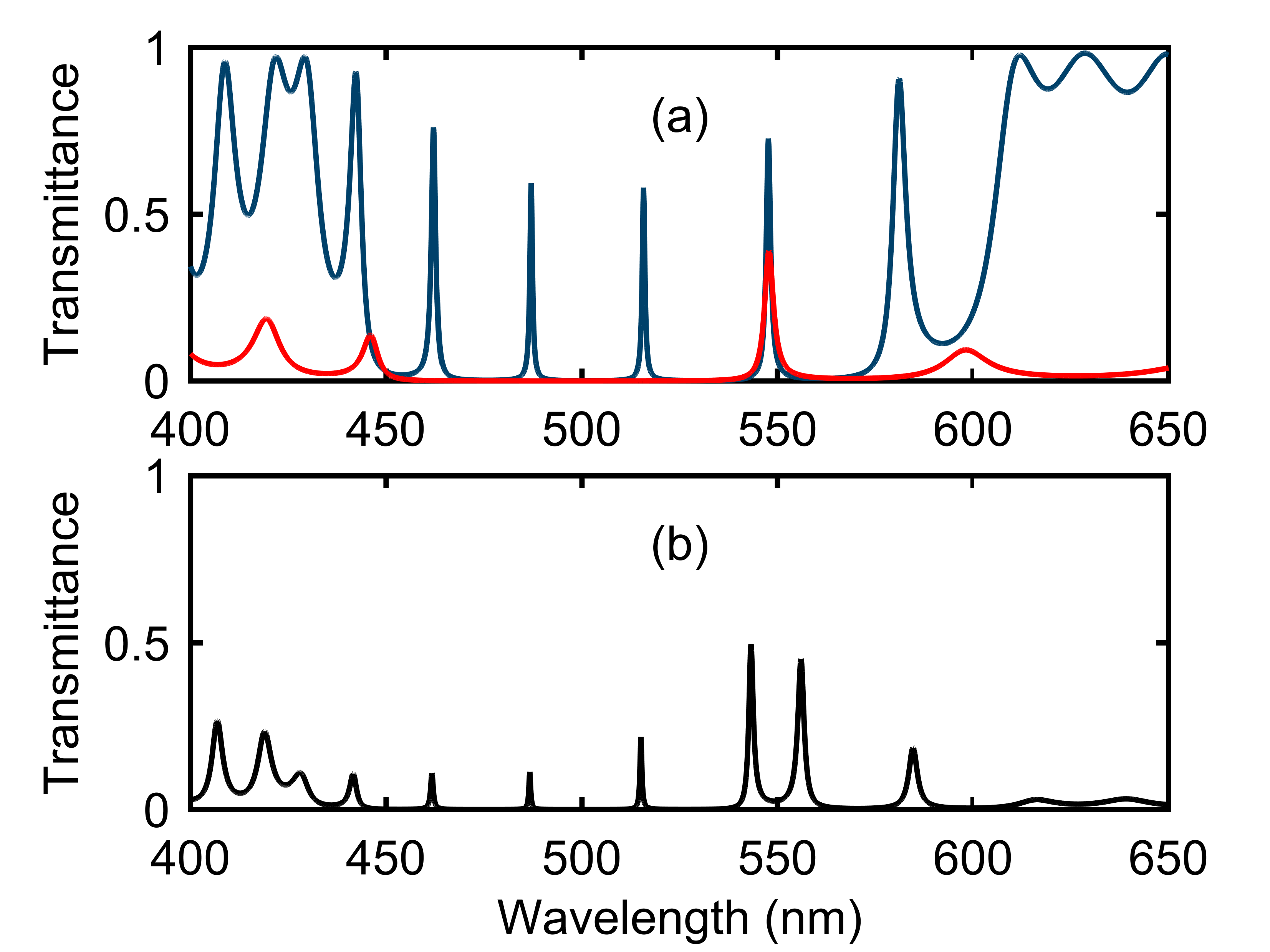}
\includegraphics[scale=1]{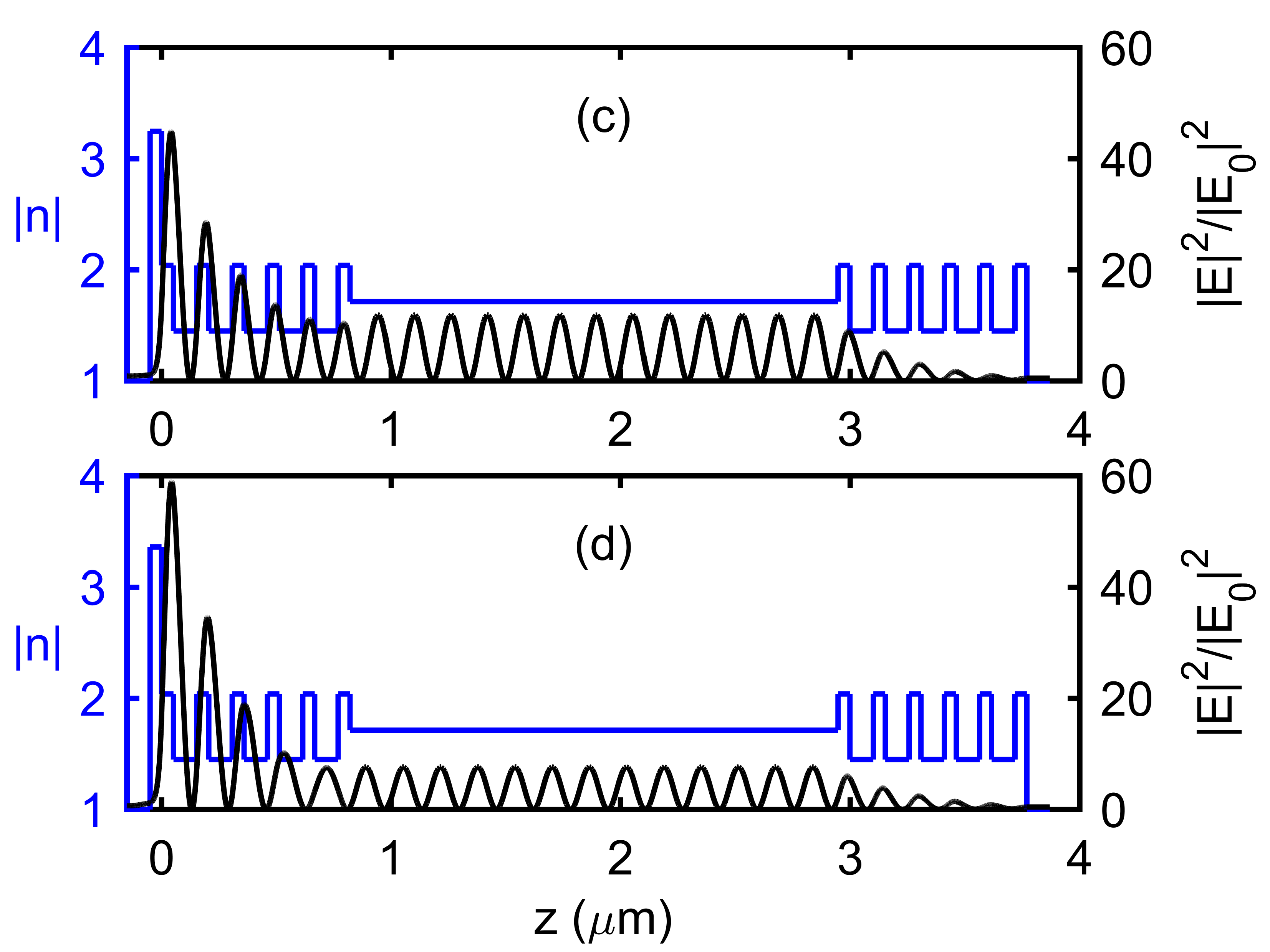}
\caption{ (a) Transmission spectra of the PCs without metallic layer (blue) (bare MC mode at 547.7 nm) and without nematic defect (red) (bare TPP at 547.7 nm). (b) Transmission spectrum of the investigated structure (hybrid Tamm-MC modes at 543.2 and 556 nm).
The nematic director is parallel to the electric field ($x$ polarization). Spatial distribution of the refractive index and electric field energy of the light wave in the hybrid Tamm-MC modes at wavelengths of (c) 543.2 and (d) 556 nm.}
\label{fig2}
\end{figure}

If we eliminate the silver layer, the structure will turn to the MC filled with the nematic bound by Bragg mirrors.
In the spectrum shown in Fig.~\ref{fig2}(a), one can see several transmission peaks localized in the PC band gap.
These peaks correspond to bare MC modes.
If there is no nematic layer in the structure, then the latter represents two conjugate mirrors: metallic and Bragg.
At the interface between them, a bare TPP is maintained.
The TPP manifests itself in the spectrum as a narrow peak (Fig.~\ref{fig2}(a)).
The parameters of the structure are found to equate the wevelengths of the TPP and a MC mode.
If the PC contains both the nematic and silver film, then both modes are simultaneously excited.
This leads to their coupling and formation of two hybrid Tamm-MC modes, short- and long-wavelength.
This can be seen in the spectrum as splitting of the transmission peak (Fig.~\ref{fig2}(b)).

The splitting value characterizes the mode coupling and, in the investigated case, is equal to 12.8~nm.
The splitting value can be increased by enhancing the spatial overlap of the coupled modes.
This can be achieved by decreasing a number of PC layers between the silver film and nematic defect.
The field energy is distributed between coupled modes localized both at the PC/metal interface and in the bulk of the MC (Fig.~\ref{fig2}(c,d)).

\subsection{Temperature control of the hybrid modes}

\begin{figure}[htbp]
\includegraphics[scale=1]{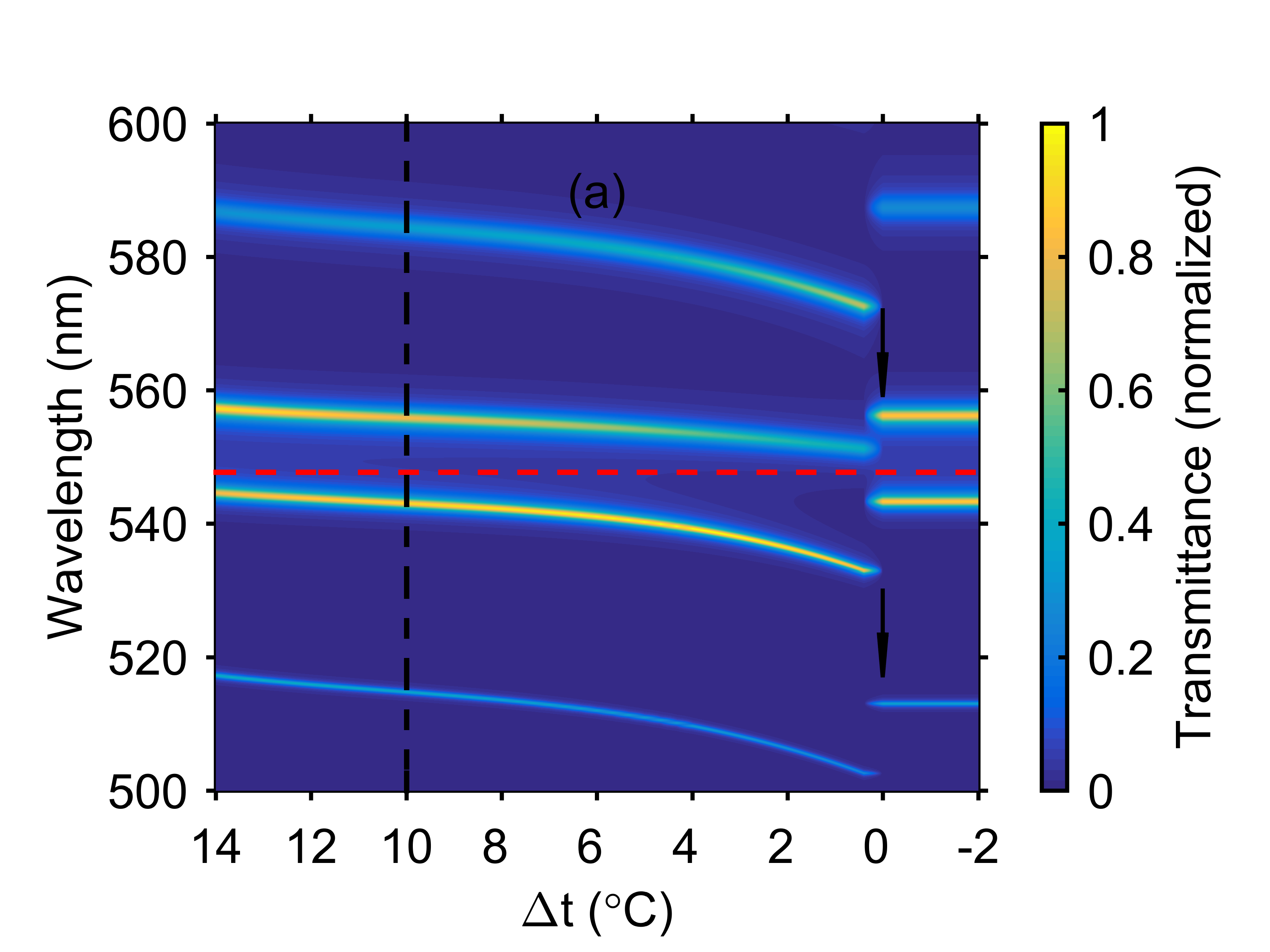}
\includegraphics[scale=1]{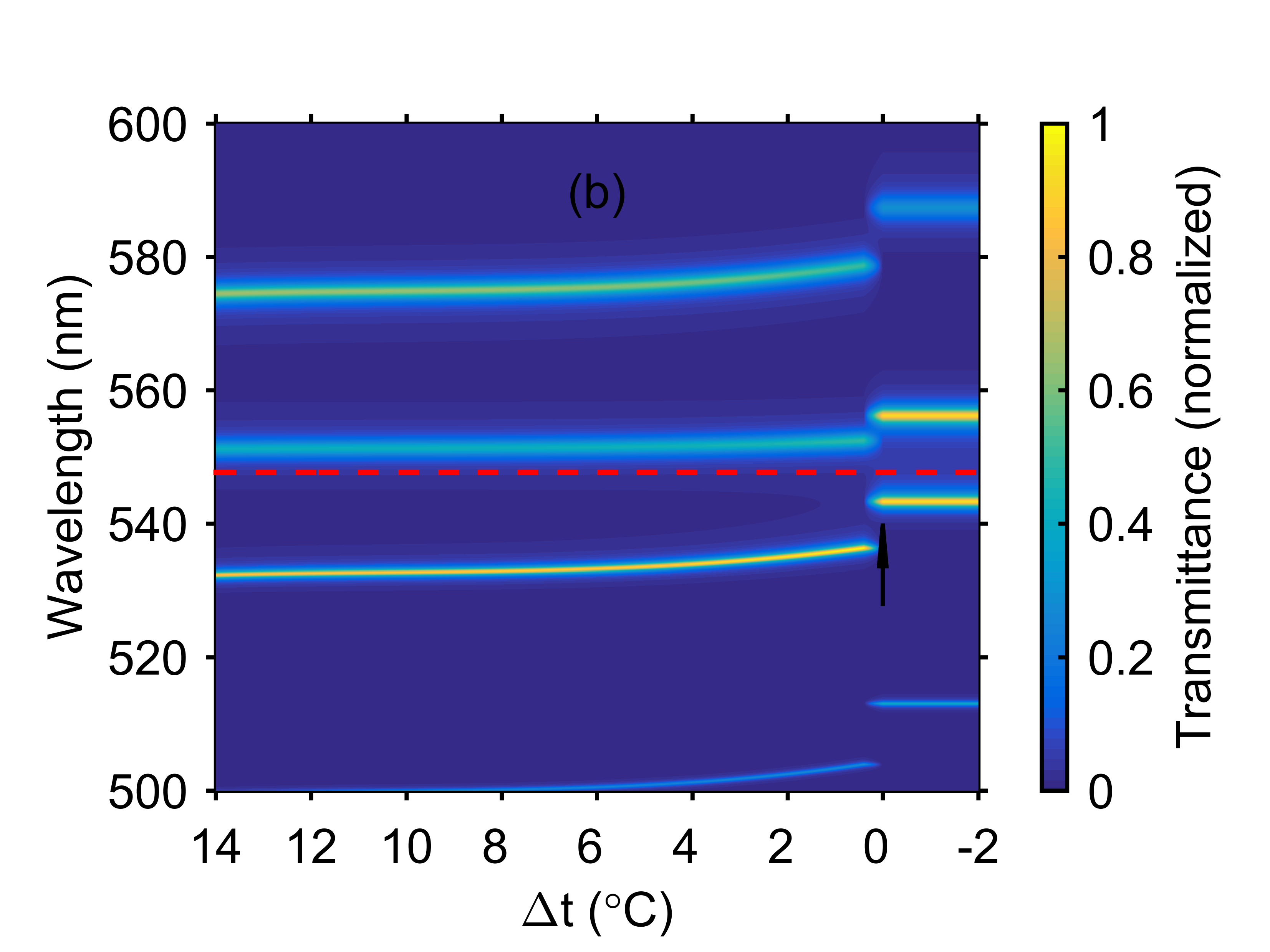}
\caption{Normalized transmission spectrum $T/T_{max}$ of the structure vs temperature detuning $\Delta t = t_0-t$, where $t_0 = 35.0^\circ$C is the phase transition temperature.
(a) The nematic director is parallel to the electric field ($x$ polarization); $T_{max}$~=~0.55.
The black line corresponds to the spectrum in Fig.~\ref{fig2}(b).
(b) The nematic director is perpendicular to the electric field vector ($y$ polarization); $T_{max}$~=~0.52.
Arrows show the direction of the jump of MC modes involved in the formation of the hybrid state.
The red line corresponds to the bare TPP wevelength.}
\label{fig3}
\end{figure}

To calculate temperature tuning of the spectra, we used the experimental values of the temperature dependences of refractive indices $n_e$ and $n_o$ ~\cite{Sefton1985}.
Figure~\ref{fig3} shows the calculated temperature transmission spectra of the structure in the wavelength range corresponding to the hybrid modes.

In the figure, one can see the motion of peaks corresponding to the hybrid modes.
Since the bare TPP wevelength remains invariable, the motion of peaks can be explained by the sensitivity of the MC mode wevelengths to the applied temperature field~\cite{Arkhipkin2008}.
As the temperature increases, the refractive index $n_e$ decreases, which leads to a decrease in the MC optical thickness and, consequently, to the blue shift of MC modes (Fig.~\ref{fig3}(a)).
It follows from the Fabry--Perot resonance condition that~\cite{Born_Wolf1999b}:

\begin{equation}
m\lambda_m=2Ln.
\label{eq7}
\end{equation}
Here, $m$ is the number of MC mode, $\lambda_m$ is its wavelength, and $n$ is the refractive index of the defect layer.
Actually the MC mode standing wave antinodes are shifted beyond the defect layer thickness $L$. The corresponding wavelength shift is neglected in Eq.~(\ref{eq7}) .

The wevelength shift of MC modes leads to the shift of hybrid Tamm-MC modes.
In the investigated temperature range, the shift of the long-wavelength mode attains 7 nm and the shift of the short-wavelength mode, 12 nm.
At the point of the phase transition between nematic and isotropic liquid at $t_0 = 35.0^\circ$C, the jump of refractive index $n_e$ is observed~\cite{Blinov2010bk}.
This leads to the stepwise blue shift of the MC modes, which attains 20 nm.
During the phase transition, the MC mode coupled with the TPP turns to the bare MC mode.
At the same time, the neighboring MC mode jumps toward the TPP; as a result, they are coupled and new bound Tamm-MC states are formed.
With a further increase in temperature, the refractive index of the isotropic liquid remains almost invariable; therefore, the hybrid mode wevelengths after the phase transition do not change.

The temperature behavior of refractive index $n_o$ (Fig.~\ref{fig3}(b)) is opposite to (Fig.~\ref{fig3}(a)). The resonant wavelength grows with temperature and the hybrid modes undergo the red shift.
Through $14^\circ$ temperature interval the long-wavelength hybrid mode shifts by 2 nm and the short-wavelength mode shifts by 5 nm.
At the phase transition point, a stepwise red shift of the MC modes of up to 10 nm is observed.
This leads to the jump of hybrid modes.
Comparison of Figs. ~\ref{fig3}(a) and ~\ref{fig3}(b) allows us to conclude that the temperature spectra are extremely sensitive to the polarization direction of the incident light.

\subsection{Electric-field control of the hybrid modes}

\begin{figure}[htbp]
\includegraphics[scale=1]{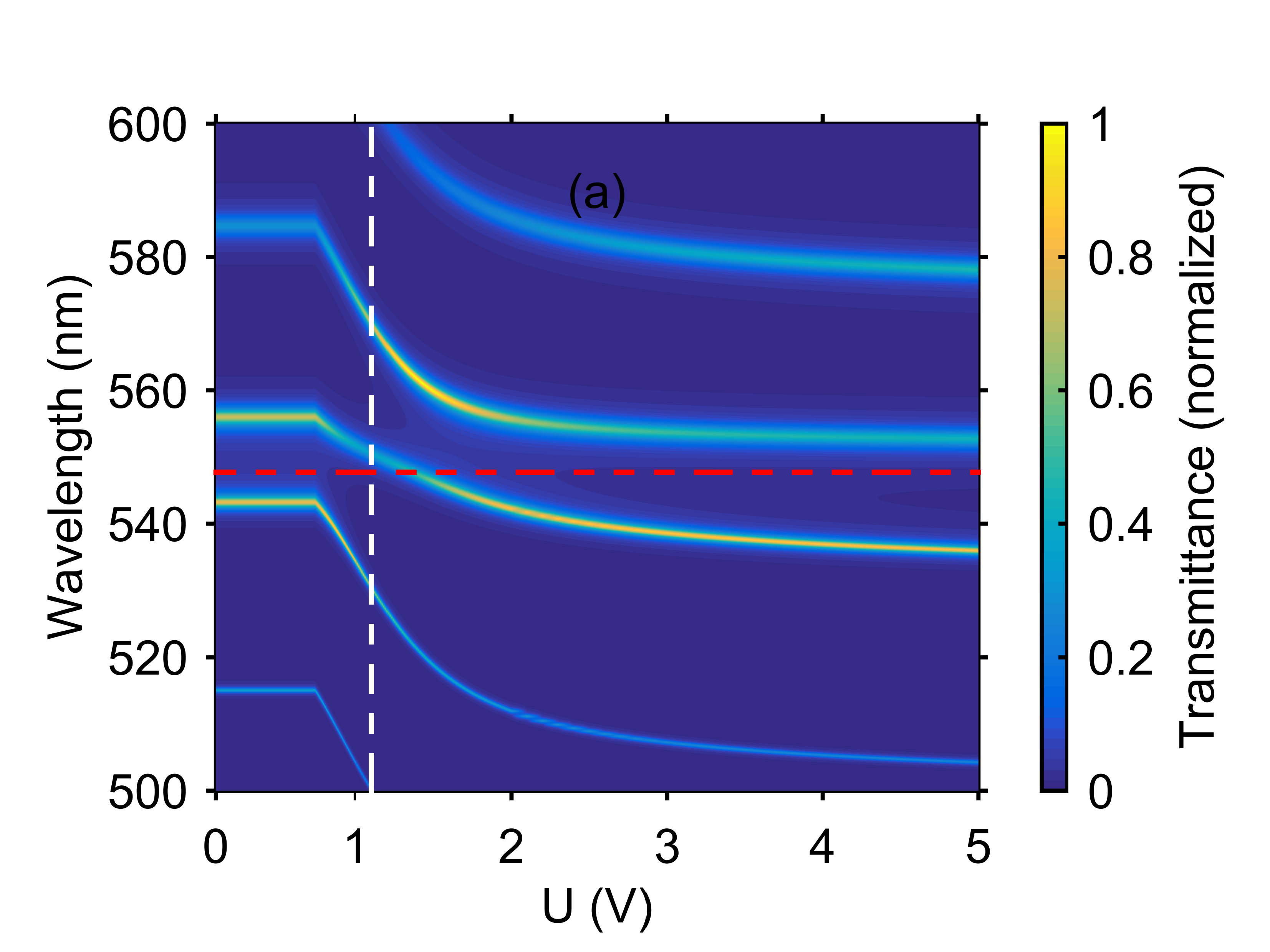}
\includegraphics[scale=1]{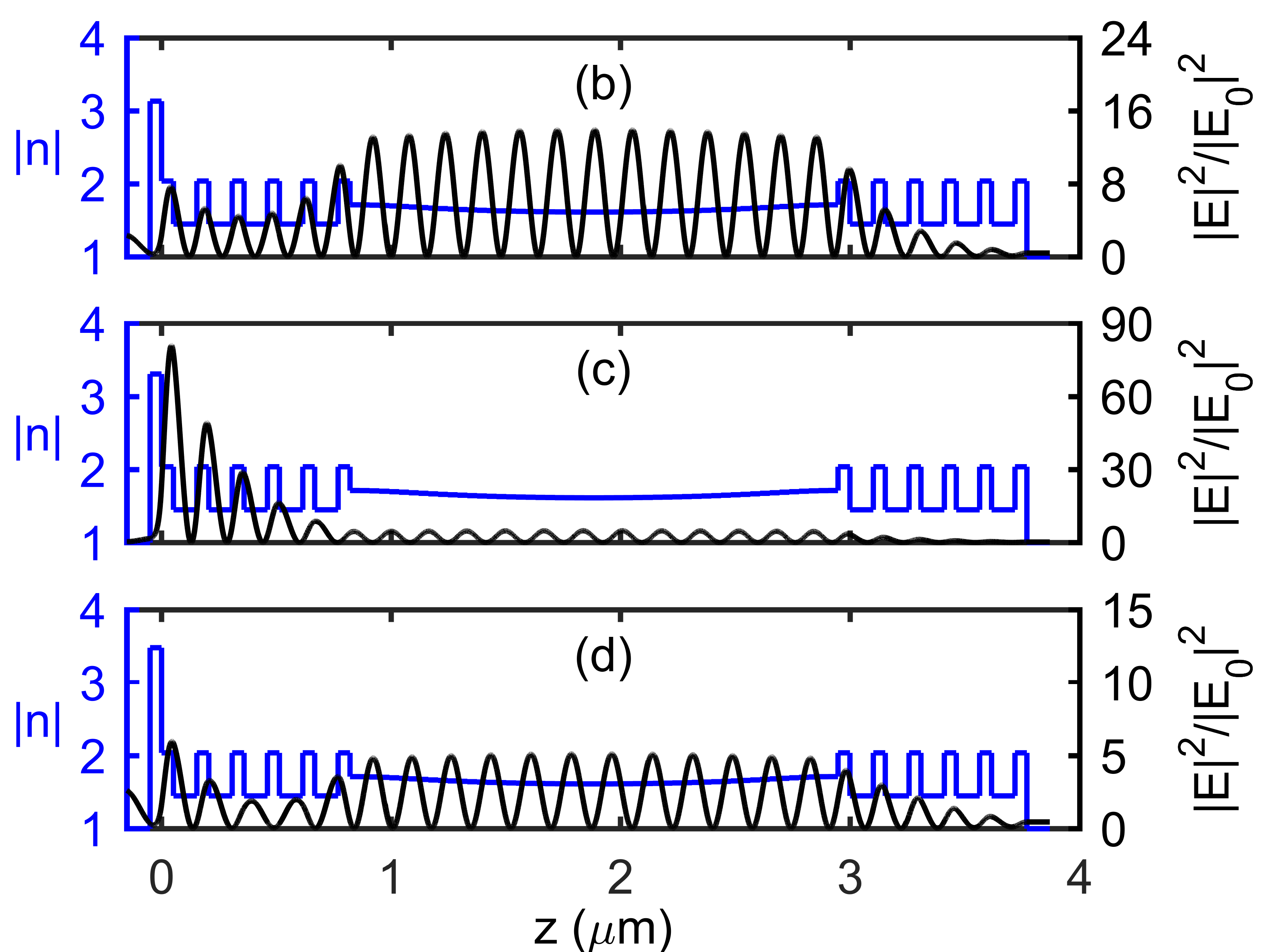}
\caption{(a) Normalized transmission spectrum $T/T_{max}$ of the structure vs applied voltage for the  $x$-polarized light; $T_{max}$~=~0.6240.
The external field is directed along the $z$ axis.
The red line corresponds to the wevelength of the bare TPP. Spatial distribution of the refractive index and electric field energy of the light wave in hybrid Tamm-MC modes at $U = 1.1$ V (white line in Fig. ~\ref{fig4}(a)) at wavelengths of (b) 530.5, (c) 550.5, and (d) 570 nm.}
\label{fig4}
\end{figure}

The electric field applied to the LC defect layer leads to the orientational transition.
This effect is called Fredericks transition ~\cite{Blinov2010bk}.
Figure~\ref{fig4}(a) shows the calculated transmission spectrum of the structure for the $x$-polarized light in the variable voltage applied to the defect layer.
It can be seen that at $U<U_c=0.74$ V the spectrum does not change.
This is caused by the threshold character of the Fredericks transition.
In this voltage range, the spectrum corresponds to Fig.~\ref{fig2}(b).

At the voltages above critical voltage $U_c$, the blue shift of hybrid modes is observed.
This can be explained as follows.
The LC in the defect layer has the initial planar orientation; the electric voltage is applied perpendicular to the defect layer.
Since the permittivity along the director is larger than the transverse component ($\varepsilon_{\vert \vert } >\varepsilon _\bot$), we observe the so-called Fredericks S-transition~\cite{Blinov2010bk}.
LC director begins to deviate from the $x$ axis and to align parallel to the applied field, i.e., along the $z$ axis.

The distribution of the tilt angle $\theta(z)$ as a function of applied voltage $U$ was found by numerical solution of Eqs. (\ref{eq5},\ref{eq6}) using the gradient descent technique~\cite{Korn_Korn2000bk,Timofeev2012}.
The boundary conditions correspond to the initial planar LC orientation $\theta(0)=\theta(L)=0$.
The calculation was performed for the following 5CB parameters at a temperature of $25^\circ$C: $k_{11} =5.9$ pN , $k_{33}= 9.9$ pN, $\varepsilon _{\vert \vert } = 18$, and $\varepsilon _\bot = 6$~\cite{Nowinowski2012}.
Taking into account the tilt of LC director, the LC refractive index for the $x$-polarized radiation is ~\cite{Born_Wolf1999b}

\begin{equation}
\label{eq8}
n(z) =\frac{n_en_o}{\sqrt{n_e^2\sin^2\theta(z)+n_o^2\cos^2\theta(z)}}.
\end{equation}

It can be seen that the refractive index of the defect LC layer decreases with increasing tilt angle.
This is shown by comparison of the refractive index distributions over the structure in Figs.\ref{fig4} and \ref{fig2}.
One can see that under the action of applied voltage, the refractive index of the nematic layer decreases over the entire LC volume and is minimum at the center of the MC where there are no boundary effects on PC mirrors.
With regard to Eq.(\ref{eq7}), this leads to the blue shift of the MC modes and, simultaneously, hybrid Tamm-MC modes.
In this case, at $U>4.5$ V, the wevelengths of hybrid modes almost stop changing.
This is due to the fact that the LC director approaches the homeotropic state and its rotation slows down.
Except for a thin surface regions, the LC director aligns along the $z$ axis.
Then, the refractive index for the $x$-polarized light becomes equal to the ordinary refractive index $n_o$.
This can be seen also from Eq.(\ref{eq8}), if we take $\theta(z)=90^\circ$.

It can be seen from Fig.~\ref{fig4}(a) that at a voltage of about $U = 1.4$ V, the hybrid mode crosses the bare TPP wevelength.
Such a behavior of the hybrid mode contradicts the well-known general theory for two modes coupling, when their wavelengths shift in opposite directions and avoid to cross each other~\cite{Haus1984bk,Rabinovich_Trubetskov1989bk}.
In our case the behavior can be explained by the triple hybridization.
As the voltage increases, one MC mode initially coupled with the TPP moves away from it and turns to the bare MC mode at high voltages.
At the same time, the neighboring MC mode approaches the TPP and is hybridized with it.
In the intermediate region, where both MC modes are sufficiently close to the TPP, three modes are hybridized.
Such a behavior of the spectral peaks was observed in~\cite{Fang2013}, where the hybrid states formed by two TPPs and one MC mode were investigated.

This assumption of triple hybridization is confirmed by the light field energy distribution over the structure at a voltage of $U = 1.1$ V [see Figs.~\ref{fig4}(b,c,d)].
It can be seen that the light field in each of the three modes has localization maxima both in the bulk of the MC and at the PC/metal interface.
The mode close to the bare TPP in wavelength (Fig.~\ref{fig4}(c)) is mainly localized at the PC/metal interface, whereas the other two modes are more like the bare TPP modes and are mainly localized in the bulk of the MC (Fig.~\ref{fig4}(b,d)).
Comparison of Figs.~\ref{fig4} and \ref{fig2} shows that we can tune not only the wevelength of hybrid modes, but also the light energy localization at the PC/metal interface, which is important for the TPP applications.
The transmission spectra of the $y$-polarized wave do not change, since in this case the refractive index of the LC remains equal to the ordinary refractive index $n_o$ during reorientation of director in the  $xz$ plane.

In addition, note that in the experiment the voltage can be applied through indium tin oxide (ITO) layers embedded in the structure on each side from the nematic~\cite{Arkhipkin2011}.
In this case, the MC mode acquires an additional phase delay and is partially absorbed in the ITO layers.
The calculation shows that at an ITO layer thickness of 30 nm, the transmittance in the peaks changes by several percent and the peaks undergo a red shift of several nanometers.

\subsection{Tuning the hybrid mode wevelengths}

\begin{figure}[htbp]
\includegraphics[scale=1]{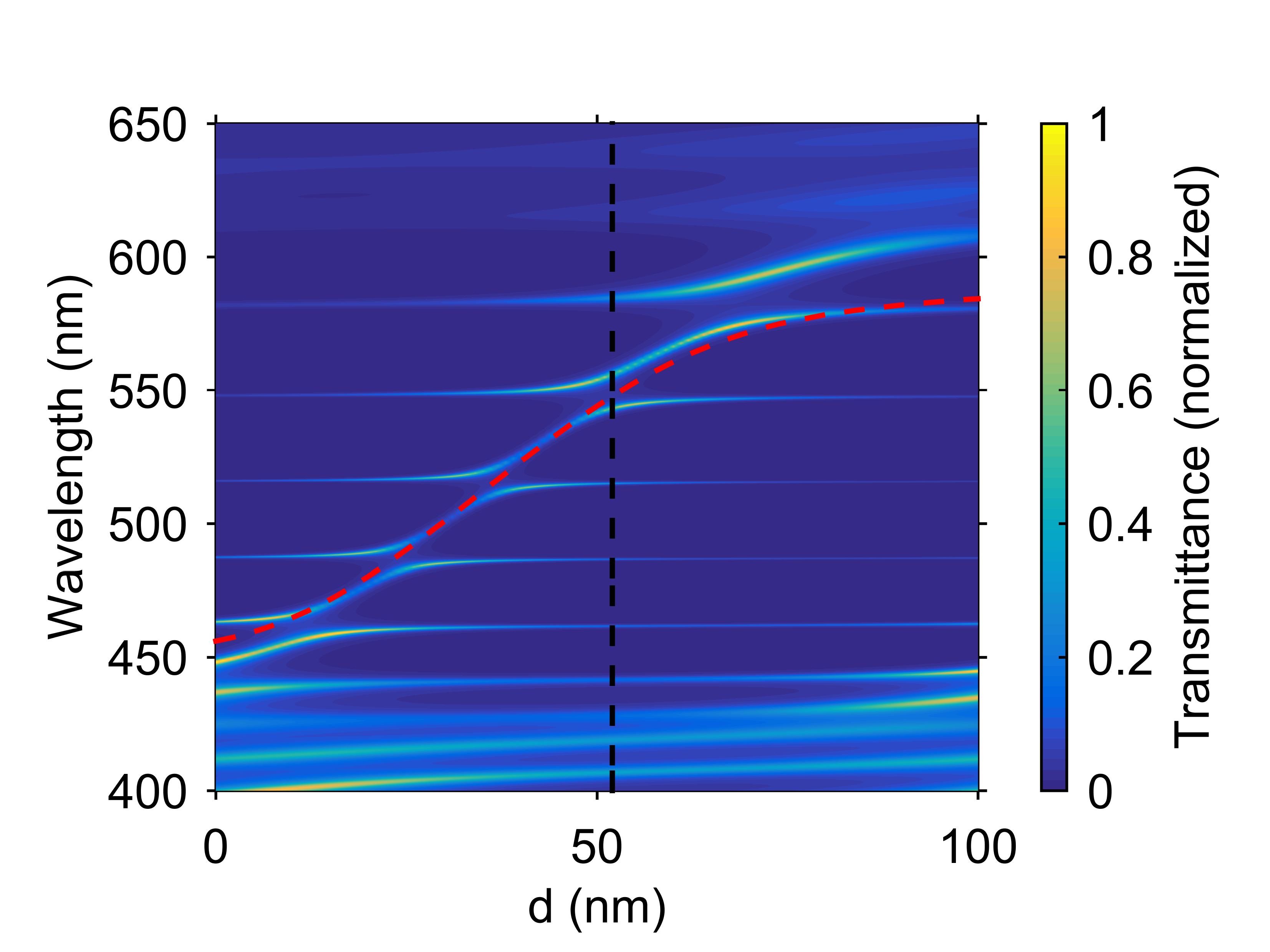}
\caption{Normalized transmission spectrum $T/T_{max}$ of the structure vs thickness $d$ of the first $ZrO_2$ layer.
The nematic director is parallel to the electric field ($x$ polarization); $T_{max}$ = 0.7186.
The black line corresponds to the spectrum in Fig.~\ref{fig2}(b).
The red line corresponds to the bare TPP wevelength.}
\label{fig5}
\end{figure}

Upon variation in thickness $d$ of the first $ZrO_2$ layer adjacent to the silver layer, the phase of reflectance from the PC changes.
Since the condition of phase matching between the reflectances from the metallic and Bragg mirrors should be satisfied, the TPP wevelength changes~\cite{Kaliteevski2007}.
This makes it possible to tune the hybrid mode wevelengths (Fig.~\ref{fig5}).

The wevelengths of the MC modes are almost insensitive to the growth of thickness $d$, while the bare TPP mode undergoes a red shift and pass through the entire PC band gap.
In this case, the sequential hybridization with all the MC modes occurs.
The similar situation is observed when the electric field vector is perpendicular to the nematic director ($y$ polarization).

The significant dependence of the transmission peak wevelengths on the first layer thickness makes it possible to design a tunable filter on the basis of the proposed structure. For this purpose, the first layer should have a variable thickness, e.g., be wedge-like~\cite{Bruckner2011}.

\section{Conclusions}

Transmission spectra of a 1D PC coated with a thin silver layer were calculated.
The photonic crystal under study contains the defect layer infiltrated with 5CB nematic.
The existence of hybrid modes formed by the Tamm plasmon polariton and microcavity mode in the structure was demonstrated.
It was shown that the light field energy is distributed over the coupled modes.

The possibility of temperature tuning of the spectra was demonstrated, which is caused by the sensitivity of the nematic refractive index to the applied temperature field.
The existence of the hybrid mode jump at the point of the phase transition nematic--isotropic liquid was demonstrated.
The polarization sensitivity of the spectra was demonstrated.

The possibility of tuning the spectra by the electric field was demonstrated, which is caused by the sensitivity of the nematic refractive index to the applied electric field.
It was shown that at the shift of the microcavity modes under the action of a field, the Tamm mode is sequentially hybridized with the neighboring microcavity modes.
It was shown that there is the voltage range where the Tamm mode is hybridized with two neighboring microcavity modes simultaneously.
The possibility of tuning the hybrid mode wevelengths by selecting a thickness of the first photonic crystal layer adjacent to the silver layer was demonstrated.

The possibility of temperature and electric-field tuning of the hybrid mode wevelengths and light energy localization at the PC surface is important for TPP applications in sensing, lasing, solar photovoltaics, WOLED technology, and nonlinear control.

\section*{Acknowledgments}

The reported study was funded by Russian Foundation for Basic Research, Government of Krasnoyarsk Territory, Krasnoyarsk Region Science and Technology Support Fund to the research project no. 17-42-240464.
P.S.P. acknowledges the support of the Scholarship of the President of the Russian Federation no. SP-227.2016.5.
The authors thank D.N.~Gulkin and Dr.~V.O.~Bessonov for help and fruitful discussions.

\bibliography{library}

\end{document}